\documentstyle[12pt]{article}

\begin{document}

\begin{center}
Quantum fluctuations and random matrix theory
\end{center}
\begin{center}
Maciej M. Duras
\end{center}
\begin{center}
Institute of Physics, Cracow University of Technology, 
ulica Podchor\c{a}\.zych 1, PL-30084 Cracow, Poland
\end{center}

\begin{center}
Email: mduras @ riad.usk.pk.edu.pl
\end{center}

\begin{center}
AD 2003 March 13
\end{center}

\begin{center}
"Fluctuations and Noise in Photonics and Quantum Optics", 
D. Abbott, J. H. Shapiro, Y. Yamamoto, Eds.; 
Proceedings of SPIE; Society of Photo-optical Instrumentation Engineers, 
Bellingham, WA, USA,  Vol. 5111, 456-459 (2003).
\end{center}

\section{Abstract}
\label{sec-Abstract}
The random matrix ensembles are applied to the quantum
statistical two-dimensional systems of electrons.
The quantum systems are studied using the finite
dimensional real, complex and quaternion Hilbert spaces of the
eigenfunctions.
The linear operators describing the systems act on these Hilbert spaces
and they are treated as random matrices in generic bases of the
eigenfunctions.
The random eigenproblems are presented and solved.
Examples of random operators
are presented with connection to physical problems.

\section{Introduction}
\label{sect:intro} 

Random Matrix Theory RMT studies
random matrix variables
\cite{Haake 1990,Guhr 1998,Mehta 1990 0,Reichl 1992,Bohigas 1991,Porter 1965,Brody 1981,Beenakker 1997}.
We study generic quantum statistical systems with energy dissipation.
Let us consider Hilbert's space $\cal{V}$ with some basis
$\{| \Psi_{i} \rangle\}$.
The space of the linear bounded operators $\hat{X}$ 
acting on Hilbert space $\cal{V}$
is called Liouville space and is denoted $\cal{L}(\cal{V})$.
The Liouville space is again Hilbert's space with the scalar product:
$\langle \hat{X} | \hat{Y} \rangle = {\rm Tr} (\hat{X}^{\dagger} \hat{Y})$.
Hence, it is a Banach space with norm:
$|| \hat{X} || = (\langle \hat{X} | \hat{X} \rangle)^{1/2}$,
and it is metric space with distance:
$\rho (\hat{X}, \hat{Y})= || \hat{X} - \hat{Y} ||$.
Finally, it is topological space with the balls
$B(\hat{X}, r)=\{ \hat{Y}| \rho (\hat{Y}, \hat{X}) < r\}$.
The Liouville space is also a differentiable manifold.
There exist a tangent space ${\rm T}_{\hat{X}}\cal{L}(\cal{V})$,
tangent bundle ${\rm T}\cal{L}(\cal{V})$,
cotangent space ${\rm T}_{\hat{X}}^{\star}\cal{L}(\cal{V})$,
and cotangent bundle ${\rm T}^{\star}\cal{L}(\cal{V})$.  
The quantum operator $\hat{X} \in \cal{L}(\cal{V})$ can be represented
in the given basis by a matrix $X \in {\rm MATRIX}(N,N,{\bf F})$ 
with matrix elements $X_{ij} \in {\bf F},$
where ${\rm MATRIX}(N,N,{\bf F})$ is set of all $N \times N$
matrices with elements from the field ${\bf F}$.
We are allowed to define random operator variable
$\hat{\cal{X}}: \Omega \ni \omega 
\rightarrow \hat{\cal{X}}(\omega) \in \cal{L}(\cal{V}),$
where $\Omega$ is sample space, $\omega$ is sample point,
and $\hat{X}=\hat{\cal{X}}(\omega)$ is deterministic operator.
It has the analogical random matrix variable
${\cal{X}}: \Omega \ni \omega 
\rightarrow {\cal{X}}(\omega) \in {\rm MATRIX}(N,N,{\bf F})$,
where $X={\cal{X}}(\omega)$ is deterministic matrix
\cite{Haake 1990,Guhr 1998,Mehta 1990 0}.
The generic nonhermitean quantum Hamiltonian operator 
$\hat{K} \in \cal{L}(\cal{V})$ 
can be represented in the given basis by a matrix 
$K \in {\rm MATRIX}(N,N,{\bf F})$ 
with matrix elements $K_{ij} \in {\bf F}$.
Thus, we define random Hamiltonian operator variable
$\hat{\cal{K}}: \Omega \ni \omega 
\rightarrow \hat{\cal{K}}(\omega) \in \cal{L}(\cal{V}),$
and we define analogical random Hamiltonian matrix variable 
${\cal{K}}: \Omega \ni \omega 
\rightarrow {\cal{K}}(\omega) \in {\rm MATRIX}(N,N,{\bf F})$.
Here, $\hat{K}=\hat{\cal{K}}(\omega)$ 
is deterministic nonhermitean Hamiltonian operator, and
$K={\cal{K}}(\omega)$ 
is deterministic nonhermitean Hamiltonian matrix.
The matrix elements
${\cal K}_{ij}$ are independent random scalar variables:
${\cal{K}}_{ij}: \Omega \ni \omega 
\rightarrow {\cal{K}}_{ij}(\omega) \in {\bf F},i,j=1, ..., N$
\cite{Haake 1990,Guhr 1998,Mehta 1990 0,Reichl 1992,Bohigas 1991,Porter 1965,Brody 1981,Beenakker 1997}.
Let us assume that the Hamiltonian operator 
$\hat{K}=\hat{\cal{K}}(\omega)$ 
is not hermitean operator, 
hence the Hamiltonian matrix $K={\cal{K}}(\omega)$ 
is not hermitean,
thus eigenenergies ${\cal{Z}}_{i}$ of ${\cal{K}}$ are
complex-valued random variables
${\cal{Z}}_{i}: \Omega \ni \omega \rightarrow {\cal{Z}}_{i}(\omega) \in {\bf F}$,
where $Z_{i}={\cal{Z}}_{i}(\omega), i=1, ..., N,$ are deterministic eigenenergies
of $K={\cal{K}}(\omega)$.
We assume that distribution of matrix elements ${\cal K}_{ij}$
is governed by complex Ginibre ensemble
\cite{Haake 1990,Guhr 1998,Ginibre 1965,Mehta 1990 1}.
$\cal{K}(\omega)$ belongs to ${\rm MATRIX}(N,N,{\bf F})$,
where ${\bf F}={\bf C}$ is complex numbers field.
Since $\hat{K}=\hat{\cal{K}}(\omega)$ is not hermitean operator,
and $K={\cal{K}}(\omega)$ is not hermitean matrix, therefore quantum system is
dissipative system. Ginibre ensemble of random matrices is one of many 
Gaussian Random Matrix ensembles GRME.
The above approach is an example of Random Matrix theory RMT
\cite{Haake 1990,Guhr 1998,Mehta 1990 0}.
The other RMT ensembles are for example
Gaussian orthogonal ensemble GOE, unitary GUE, symplectic GSE,
as well as circular ensembles: orthogonal COE,
unitary CUE, and symplectic CSE.

There were studied among others the following
Gaussian Random Matrix ensembles GRME:
orthogonal GOE, unitary GUE, symplectic GSE,
as well as circular ensembles: orthogonal COE,
unitary CUE, and symplectic CSE.
The choice of ensemble is based on quantum symmetries
ascribed to the hermitean Hamiltonian matrix $H$. 
The hermitean Hamiltonian matrix $H$
acts on space $W$ of eigenvectors.
It is assumed that $W$ is $N$-dimensional Hilbert space
$W={\bf F}^{N}$, where the real, complex, or quaternion
field ${\bf F}={\bf R, C, H}$,
corresponds to GOE, GUE, or GSE, respectively.
If the Hamiltonian matrix $H$ is hermitean $H=H^{\dag}$,
then the probability density function of random Hamiltonian matrix 
${\cal H}$ reads:
\begin{eqnarray}
& & f_{{\cal H}}(H)={\cal C}_{H \beta} 
\exp{[-\beta \cdot \frac{1}{2} \cdot {\rm Tr} (H^{2})]},
\label{pdf-GOE-GUE-GSE} \\
& & {\cal C}_{H \beta}=(\frac{\beta}{2 \pi})^{{\cal N}_{H \beta}/2}, 
\nonumber \\
& & {\cal N}_{H \beta}=N+\frac{1}{2}N(N-1)\beta, \nonumber \\
& & \int f_{{\cal H}}(H) dH=1,
\nonumber \\
& & dH=\prod_{i=1}^{N} \prod_{j \geq i}^{N} 
\prod_{\gamma=0}^{D-1} dH_{ij}^{(\gamma)}, \nonumber \\
& & H_{ij}=(H_{ij}^{(0)}, ..., H_{ij}^{(D-1)}) \in {\bf F}, \nonumber
\end{eqnarray}
where the parameter $\beta$ assume values 
$\beta=1,2,4,$  for GOE($N$), GUE($N$), GSE($N$), respectively,
and ${\cal N}_{H \beta}$ is number of independent matrix elements
of hermitean Hamiltonian matrix $H$.
The Hamiltonian matrix $H$ is hermitean $N \times N {\bf F}$-matrix,
and the matrix Haar's measure $dH$ is invariant under
transformations from the unitary group U($N$, {\bf F}).
The eigenenergies 
${\cal{E}}_{i}: \Omega \ni \omega \rightarrow {\cal{E}}_{i}(\omega) \in {\bf R}$ 
of ${\cal {H}}$ are real-valued random variables, 
where $E_{i}={\cal{E}}_{i}(\omega)$ are deterministic eigenergies of $H$, 
and $E_{i}=E_{i}^{\star}, i=1, ..., N$.
It was Eugene Wigner who firstly dealt with eigenenergy level repulsion
phenomenon studying nuclear spectra \cite{Haake 1990,Guhr 1998,Mehta 1990 0}.
RMT is applicable now in many branches of physics:
nuclear physics (slow neutron resonances, highly excited complex nuclei),
condensed phase physics (fine metallic particles,  
random Ising model [spin glasses]),
quantum chaos (quantum billiards, quantum dots), 
disordered mesoscopic systems (transport phenomena),
quantum chromodynamics, quantum gravity, field theory.

\section{The Ginibre ensembles}
\label{sec-ginibre-ensembles}

Jean Ginibre considered another example of GRME
dropping the assumption of hermiticity of Hamiltonians
thus defining generic ${\bf F}$-valued Hamiltonian $K$
\cite{Haake 1990,Guhr 1998,Ginibre 1965,Mehta 1990 1}.
Hence, $K$ belongs to ${\rm MATRIX}(N,N,{\bf F})$,
and the matrix Haar's measure $dK$ is invariant under
transformations form the general linear Lie group GL($N$, {\bf F}).
The distribution of random nonhermitean Hamiltonian variable ${\cal K}$ 
is given by:
\begin{eqnarray}
& & f_{{\cal K}}(K)={\cal C}_{K \beta} 
\exp{[-\beta \cdot \frac{1}{2} \cdot {\rm Tr} (K^{\dag}K)]},
\label{pdf-Ginibre} \\
& & {\cal C}_{K \beta}=(\frac{\beta}{2 \pi})^{{\cal N}_{K \beta}/2}, 
\nonumber \\
& & {\cal N}_{K \beta}=N^{2}\beta, \nonumber \\
& & \int f_{{\cal K}}(K) dK=1,
\nonumber \\
& & dK=\prod_{i=1}^{N} \prod_{j=1}^{N} 
\prod_{\gamma=0}^{D-1} dK_{ij}^{(\gamma)}, \nonumber \\
& & K_{ij}=(K_{ij}^{(0)}, ..., K_{ij}^{(D-1)}) \in {\bf F}, \nonumber
\end{eqnarray}
where $\beta=1,2,4$, stands for real, complex, and quaternion
Ginibre ensembles, respectively.
Therefore, the eigenenergies $Z_{i}$ of quantum system 
ascribed to Ginibre ensemble are complex-valued random variables.
The eigenenergies $Z_{i}, i=1, ..., N$,
of nonhermitean Hamiltonian $K$ are not real-valued random variables
$Z_{i} \neq Z_{i}^{\star}$.
Jean Ginibre postulated the following
joint probability density function 
of random vector of complex eigenvalues 
${\cal{Z}}_{1}, ..., {\cal{Z}}_{N}$
for $N \times N$ Hamiltonian matrices $K$ for $\beta=2$
\cite{Haake 1990,Guhr 1998,Ginibre 1965,Mehta 1990 1}:
\begin{equation}
P(Z_{1}, ..., Z_{N})= 
\prod _{j=1}^{N} \frac{1}{\pi \cdot j!} \cdot
\prod _{i<j}^{N} \vert Z_{i} - Z_{j} \vert^{2} \cdot
\exp (- \sum _{j=1}^{N} \vert Z_{j}\vert^{2}),
\label{Ginibre-joint-pdf-eigenvalues} 
\end{equation}
where $Z_{i}$ are complex-valued sample points
($Z_{i} \in {\bf C}$).
 
We emphasize here Wigner and Dyson's electrostatic analogy.
A Coulomb gas of $N$ unit charges moving on complex plane (Gauss's plane)
{\bf C} is considered. The vectors of positions
of charges are $Z_{i}$ and potential energy of the system is:
\begin{equation}
U(Z_{1}, ...,Z_{N})=
- \sum_{i<j} \ln \vert Z_{i} - Z_{j} \vert
+ \frac{1}{2} \sum_{i} \vert Z_{i}^{2} \vert. 
\label{Coulomb-potential-energy}
\end{equation}
If gas is in thermodynamical equilibrium at temperature
$T= \frac{1}{2 k_{B}}$ 
($\beta= \frac{1}{k_{B}T}=2$, $k_{B}$ is Boltzmann's constant),
then probability density function of vectors of positions is 
$P(Z_{1}, ..., Z_{N})$ Eq. (\ref{Ginibre-joint-pdf-eigenvalues}).
Therefore, complex eigenenergies $Z_{i}$ of quantum system 
are analogous to vectors of positions of charges of Coulomb gas.
Moreover, complex-valued spacings $\Delta^{1} Z_{i}$
of complex eigenenergies of quantum system:
\begin{equation}
\Delta^{1} Z_{i}=Z_{i+1}-Z_{i}, i=1, ..., (N-1),
\label{first-diff-def}
\end{equation}
are analogous to vectors of relative positions of electric charges.
Finally, complex-valued
second differences $\Delta^{2} Z_{i}$ of complex eigenenergies:
\begin{equation}
\Delta ^{2} Z_{i}=Z_{i+2} - 2Z_{i+1} + Z_{i}, i=1, ..., (N-2),
\label{Ginibre-second-difference-def}
\end{equation}
are analogous to
vectors of relative positions of vectors
of relative positions of electric charges.

The eigenenergies $Z_{i}=Z(i)$ can be treated as values of function $Z$
of discrete parameter $i=1, ..., N$.
The "Jacobian" of $Z_{i}$ reads:
\begin{equation}
{\rm Jac} Z_{i}= \frac{\partial Z_{i}}{\partial i}
\simeq \frac{\Delta^{1} Z_{i}}{\Delta^{1} i}=\Delta^{1} Z_{i}.
\label{jacobian-Z}
\end{equation}
We readily have, that the spacing is an discrete analog of Jacobian,
since the indexing parameter $i$ belongs to discrete space
of indices $i \in I=\{1, ..., N \}$. Therefore, the first derivative
with respect to $i$ reduces to the first differential quotient.
The Hessian is a Jacobian applied to Jacobian.
We immediately have the formula for the diagonal element 
of discrete "Hessian" for the eigenenergies $Z_{i}$:
\begin{equation}
{\rm Hess} Z_{i}= \frac{\partial ^{2} Z_{i}}{\partial i^{2}}
\simeq \frac{\Delta^{2} Z_{i}}{\Delta^{1} i^{2}}=\Delta^{2} Z_{i}.
\label{hessian-Z}
\end{equation}
Thus, the second difference of $Z$ is discrete analog of Hessian of $Z$.
One emphasizes that both "Jacobian" and "Hessian"
work on discrete index space $I$ of indices $i$.
The spacing is also a discrete analog of energy slope
whereas the second difference corresponds to
energy curvature with respect to external parameter $\lambda$
describing parametric ``evolution'' of energy levels
\cite{Zakrzewski 1,Zakrzewski 2}.
The finite differences of order higher than two
are discrete analogs of compositions of "Jacobians" with "Hessians" of $Z$.

The eigenenergies $E_{i}, i \in I$, of the hermitean Hamiltonian $H$
are ordered increasingly real-valued random variables.
They are values of discrete function $E_{i}=E(i)$.
The first difference of adjacent eigenenergies is:
\begin{equation}
\Delta^{1} E_{i}=E_{i+1}-E_{i}, i=1, ..., (N-1),
\label{first-diff-def-GRME}
\end{equation}
are analogous to vectors of relative positions of electric charges
of one-dimensional Coulomb gas. It is simply the spacing of two adjacent
energies.
Real-valued
second differences $\Delta^{2} E_{i}$ of eigenenergies:
\begin{equation}
\Delta ^{2} E_{i}=E_{i+2} - 2E_{i+1} + E_{i}, i=1, ..., (N-2),
\label{Ginibre-second-difference-def-GRME}
\end{equation}
are analogous to vectors of relative positions 
of vectors of relative positions of charges of one-dimensional
Coulomb gas.
The $\Delta ^{2} Z_{i}$ have their real parts
${\rm Re} \Delta ^{2} Z_{i}$,
and imaginary parts
${\rm Im} \Delta ^{2} Z_{i}$, 
as well as radii (moduli)
$\vert \Delta ^{2} Z_{i} \vert$,
and main arguments (angles) ${\rm Arg} \Delta ^{2} Z_{i}$.
$\Delta ^{2} Z_{i}$ are extensions of real-valued second differences:
\begin{equation}
\Delta^{2} E_{i}=E_{i+2}-2E_{i+1}+E_{i}, i=1, ..., (N-2),
\label{second-diff-def}
\end{equation}
of adjacent ordered increasingly real-valued eigenenergies $E_{i}$
of Hamiltonian $H$ defined for
GOE, GUE, GSE, and Poisson ensemble PE
(where Poisson ensemble is composed of uncorrelated
randomly distributed eigenenergies)
\cite{Duras 1996 PRE,Duras 1996 thesis,Duras 1999 Phys,Duras 1999 Nap,Duras 2000 JOptB}.
The Jacobian and Hessian operators of energy function $E(i)=E_{i}$
for these ensembles read:
\begin{equation}
{\rm Jac} E_{i}= \frac{\partial E_{i}}{\partial i}
\simeq \frac{\Delta^{1} E_{i}}{\Delta^{1} i}=\Delta^{1} E_{i},
\label{jacobian-E}
\end{equation}
and
\begin{equation}
{\rm Hess} E_{i}= \frac{\partial ^{2} E_{i}}{\partial i^{2}}
\simeq \frac{\Delta^{2} E_{i}}{\Delta^{1} i^{2}}=\Delta^{2} E_{i}.
\label{hessian-E}
\end{equation}
The treatment of first and second differences of eigenenergies
as discrete analogs of Jacobians and Hessians
allows one to consider these eigenenergies as a magnitudes 
with statistical properties studied in discrete space of indices.
The labelling index $i$ of the eigenenergies is
an additional variable of "motion", hence the space of indices $I$
augments the space of dynamics of random magnitudes.
 
%%%%%%%%%%%%%%%%%%%%%%%%%%%%%%%%%%%%%%%%%%%%%%%%%%%%%%%%%%%%%
%%%%% References %%%%%

\end{document}